\documentclass[12pt,a4paper]{article}
\usepackage{amssymb}

\begin{document}

\newcommand{\be}{\begin{equation}}
\newcommand{\ee}{\end{equation}}
\newcommand{\beann}{\begin{eqnarray*}}
\newcommand{\eeann}{\end{eqnarray*}}
\newcommand{\bea}{\begin{eqnarray}}
\newcommand{\eea}{\end{eqnarray}}
\newcommand{\lb}{\label}
\newcommand{\bdm}{\begin{displaymath}}
\newcommand{\edm}{\end{displaymath}}
\newcommand{\D}{{\rm d}}
\newcommand{\E}{{\rm e}}
\newcommand{\I}{{\rm i}}

\begin{titlepage}

\noindent

\vspace*{1cm}
\begin{center}
{\large\bf HAWKING TEMPERATURE FROM QUASI-NORMAL MODES} 

\vskip 1cm

{\bf Claus Kiefer} 
\vskip 0.4cm
Institut f\"ur Theoretische Physik,\\ Universit\"{a}t zu K\"oln, \\
Z\"ulpicher Str.~77,
50937 K\"oln, Germany.\\
\vspace{1cm}

\nopagebreak[4]

\begin{abstract}
A perturbed black hole has characteristic frequencies (quasi-normal modes).
Here I apply a quantum measurement analysis of the quasi-normal mode
frequency in the limit of high damping. It turns out that a measurement
of this mode necessarily adds noise to it. For a Schwarzschild
black hole, this corresponds exactly to the Hawking
temperature. The situation for other black holes is briefly discussed.
\end{abstract}
\end{center}

\end{titlepage}

Stationary black holes in four spacetime dimensions are characterized
by just three numbers: mass, angular momentum, and electric charge.
If a black hole is perturbed, however, it exhibits characteristic
oscillations called quasi-normal modes (QNMs) because they are
damped in time (and thus have a complex frequency). They correspond to
perturbations of the geometry (and other fields, if present) outside the
horizon and obey the boundary conditions of being ingoing at the horizon
and outgoing at infinity, see for example \cite{reviews} for reviews.
I shall restrict myself here to purely gravitational perturbations.

Consider in the following a Schwarzschild black hole, which is
fully characterized by its mass, $M$. Two limits 
for the quasi-normal modes are of particular interest. First, if the
angular momentum, $\ell$, of the perturbation goes to infinity, one has
for the QNM frequency (setting $c=1$),
\be
\lb{ltoinfinity}
\omega_{\ell}\sim \frac{\ell +\frac{1}{2}}{3\sqrt{3}GM}
 -\I\frac{\sqrt{3}\left(n+\frac{1}{2}\right)}{9GM}\ ,
\ee
where $n\in{\mathbb N}_0$.
The real part of $\omega_{\ell}$ is much bigger than the imaginary part.
These are the modes that are also of importance for the form of the
gravitational waves searched for in present interferometers.
Second, for fixed $\ell$ there exist,
due to the boundary conditions, a countable infinite number of modes
(labelled by $n\in{\mathbb N}_0$), which in the limit $n\to\infty$
obey
\be
\lb{omegan}
\omega_n=\frac{\ln 3}{8\pi GM}-\I\frac{n+\frac{1}{2}}{4GM}
+{\mathcal O}\left(n^{-1/2}\right) \ .
\ee
For these modes, the imaginary part is much bigger than the real part;
they are thus highly damped. In spite of this, they are of central
importance, for the following reasons. First, the imaginary parts
of the $\omega_n$ are equidistantly spaced and could thus be of relevance
for `Euclidean quantum gravity', where imaginary time (and thus frequency)
is related to temperature, cf. \cite{oup} for an introduction into
approaches to quantum gravity. 
Second, the real part of $\omega_n$,
\be
\lb{realpart}
{\rm Re}\, \omega_n\equiv\omega_{\rm QNM}=\frac{\ln 3}{8\pi GM}
\approx 8.85\frac{M_{\odot}}{M}\ {\rm kHz}\ ,
\ee
is {\em universal} in the sense that it is independent of $\ell$ and
$n$ and thus only dependent on the mass. It can be considered as a
property of the black hole itself.

Introducing the Hawking temperature,
\be
\lb{TH}
T_{\rm H}=\frac{\hbar}{8\pi k_{\rm B}GM}\ ,
\ee
one can write (\ref{omegan}) in the form,
\be
\lb{omegan2}
\hbar\omega_n=k_{\rm B}T_{\rm H}\left(\ln 3-2\pi\I\left[n+\frac{1}{2}\right]
 \right)+{\mathcal O}\left(n^{-1/2}\right)\ .
\ee
This suggests a possible connection to the quantum features of the black hole.
In fact, it has been proposed long ago that the area of a black hole
should be quantized \cite{BM}. 
Equal spacing for the area would lead to a spacing in mass,
\be
\lb{DeltaM}
\Delta M=\frac{\hbar\ln k}{8\pi GM}\equiv\hbar\tilde{\omega}_k\ ,
\ \ k=2,3,\ldots\ ,
\ee
where $\tilde{\omega}_k$ would be the `emission frequency'
(treating a black hole in analogy to an atom). 
Identification of this frequency with the QNM frequency (\ref{realpart})
would fix $k=3$. (Actually, the idea to identify these
frequencies led to the suggestion that the factor $\ln 3$ should
appear in (\ref{realpart}) \cite{Hod}. This was then proven in \cite{Motl},
see also \cite{MN}.) 
On the other hand, a discrete area spectrum, albeit not equidistant,
 was found in loop quantum gravity,
see for example \cite{oup} and the references therein. The requirement
that the black-hole entropy from counting states in loop quantum gravity
leads to the Bekenstein--Hawking entropy, together with the identification
of $\Delta M/\hbar$ corresponding to the minimal area change with  
(\ref{realpart}), fixes an ambiguity (the `Barbero--Immirzi parameter')
present in that approach (provided the gauge group is SO(3)) \cite{Dreyer}.

\medskip

It has not yet been understood why the classical frequency (\ref{realpart})
is so closely related to the Hawking temperature (\ref{TH}), which is
a quantum effect. In the following I will show that (\ref{TH}) necessarily
follows from a quantum measurement analysis of the QNM frequency.
The idea behind this is the fact that also the perturbations described
by the QNMs have to be described fundamentally by quantum theory. 
Attempting to describe a black hole as a harmonic oscillator with
frequency (\ref{ltoinfinity}) or (\ref{omegan}) is possible, although
the hole is a very poor oscillator, the `quality factor' (ratio of
real part to imaginary part of the frequency) being very small,
cf. \cite{LM}.\footnote{The quality factor could increase after the
back reaction of the Hawking radiation on the metric is taken into
account, cf. \cite{Ko}.} 

The measurement of oscillator position (here: of field amplitude)
is, of course, limited by the uncertainty relation. A detailed analysis
exhibits, however, a stronger limitation \cite{Caves}:
In order to process a quantum signal, an `amplifier' must be coupled
to the mode to be measured. This necessarily adds noise to the signal.
(Interestingly, this was first discussed in the context of 
gravitational-wave measurements, where it is still of relevance
\cite{GW}.)
In the simplest case one considers a single-mode input--output
and a linear amplifier. The amplification process can then be
described by \cite{Caves}
\be
\lb{amp}
a_{\rm out}=Ma_{\rm in}+La_{\rm in}^{\dagger}+{\mathcal F}\ ,
\ee
where $a_{\rm in}$, $a_{\rm out}$, ${\mathcal F}$ denote
the annihilation operators for the input signal, the output signal, and
the added noise, respectively.
 The coefficients obey the relation (coming from
the condition of unitarity),
\be
\lb{unitarity}
\vert M\vert^2-\vert L\vert^2-[{\mathcal F},{\mathcal F}^{\dagger}]=1 \ .
\ee
In the following I restrict myself to `phase-insensitive amplifiers'. 
They are defined by the fact that $\langle a_{\rm out}\rangle$
is invariant under arbitrary phase transformations. One can show that this
entails either $L=0$ (`phase preserving') or $M=0$ (`phase conjugating').
Introducing for an operator $R=R_1+\I R_2$ the mean-square fluctuation,
\bdm
\vert\Delta R\vert^2=\frac{1}{2}\langle RR^{\dagger}+R^{\dagger}R\rangle
-\langle R\rangle\langle R^{\dagger}\rangle=(\Delta R_1)^2+(\Delta R_2)^2\ ,
\edm
one can define the `gain' (measured in `numbers of quanta'),
of an amplifier by ${\mathcal G}=\vert M\vert^2+\vert L\vert^2$,
leading to
\be
\lb{gain}
\vert\Delta a_{\rm out}\vert^2={\mathcal G}\vert\Delta a_{\rm in}\vert^2
+\vert\Delta{\mathcal F}\vert^2\ .
\ee
The `added noise number', $A$, is then defined by
\be
\lb{noise}
A\equiv \frac{\vert\Delta{\mathcal F}\vert^2}{\mathcal G}\ .
\ee
Making in particular use of the unitarity condition (\ref{unitarity}),
Caves proves the `fundamental theorem' for the added noise \cite{Caves},
\be
\lb{theorem}
A\geq \frac{1}{2}\left\vert1\mp {\mathcal G}^{-1}\right\vert\ ,
\ee
where the upper (lower) sign corresponds to phase preserving
(phase conjugating).
A high-gain (${\mathcal G}>1$) phase-insensitive amplifier thus
necessarily adds noise to the signal.\footnote{The generalization to
phase-sensitive amplifiers can be made and leads to an uncertainty relation
for the noises connected with different phases \cite{Caves}.}

Sometimes it is useful to introduce the concept of a `noise
temperature', $T_n$. This is defined as the increase in input temperature
that is needed to account for the output noise. In the simplest case
of vanishing input temperature, $T_n$ is defined by
\be
\lb{temp}
\frac{1}{2}+\frac{1}{\E^{\hbar\omega_{\rm in}/k_{\rm B}T_n}-1}
\equiv\frac{1}{2}{\rm coth}\frac{\hbar\omega_{\rm in}}{2k_{\rm B}T_n}
\equiv A+\frac{1}{2}\ ,
\ee
where $\omega_{\rm in}$ denotes the frequency of the input mode.
The first term on the left-hand side just denotes the mean energy
(in units of $\hbar\omega_{\rm in}$) of a harmonic oscillator
at a temperature $T_n$.
One recognizes that $T_n=0$ corresponds to $A=0$. The added noise number 
has thus been rewritten in a form where it mimics an added
temperature, at least with respect to certain expectation values.

 From the theorem (\ref{theorem}), one then gets a lower limit on the noise
temperature,\footnote{This is of relevance for experimental approaches
to reach the quantum limit, cf. the recent experiments described
in \cite{exp}.}
\be
\lb{limit}
T_n\geq \frac{\hbar\omega_{\rm in}}{k_{\rm B}}
\ln^{-1}\left(\frac{3\mp{\mathcal G}^{-1}}{1\mp{\mathcal G}^{-1}}\right)\;\;
\stackrel{{\mathcal G}\to\infty}{\longrightarrow}\;\;
\frac{\hbar\omega_{\rm in}}
{k_{\rm B}\ln 3}\ .
\ee
Applying this to the QNM frequency of a Schwarzschild black hole,
one has to use the limit of infinite gain, because the signal is
highly damped. Choosing $\omega_{\rm in}=\omega_{\rm QNM}$ and
inserting (\ref{realpart}) into (\ref{limit}), one obtains
\be
\lb{result}
T_n\geq T_{\rm H}\ ,
\ee
that is, the minimal noise temperature is just given by the
Hawking temperature (\ref{TH})! In other words, 
one could thus have introduced the
Hawking temperature as the temperature  that gives the lower limit
to the noise that is added through a quantum measurement of the 
highly-damped quasi-normal modes. 

What about the situation for other types of black holes? 
For Schwarzschild black holes in dimensions higher than four,
one still has $\hbar\omega_{\rm QNM}=(\ln 3)k_{\rm B}T_{\rm H}$ 
\cite{MN, Kunst} and 
(\ref{result}) thus still holds. For the Reissner--Nordstr\"om black hole,
however, the situation is more complicated already in four spacetime
dimensions \cite{MN,AH}: There is no equal spacing between the
imaginary parts of the frequency, and $\omega_{\rm QNM}$ is not related
to the Hawking temperature in a simple way. Still, of course, 
the quantum measurement analysis will introduce a noise temperature
which could be bounded by $T_{\rm H}$. Interestingly, in the limit
of an extremal black hole (where the charge $Q$ obeys
$Q=\sqrt{G}M$), $\omega_{\rm QNM}$ is given by the Schwarzschild
expression (\ref{realpart}). This necessarily gives the same bound
as (\ref{result}), that is, the noise temperature is again 
bounded by the Hawking
temperature for a {\em Schwarzschild} black hole
(the Hawking temperature of the extremal Reissner--Nordstr\"om
black hole is zero). For a Kerr black hole it has been found that
$\omega_{\rm QNM}=mf(a)$, with an unknown function of the rotation
parameter $a$ ($m$ denotes the azimuthal angular momentum) \cite{BCY}.
 The frequency seems to be no simple polynomial
function of the Hawking temperature and the angular velocity, but
it is important that the QNM frequency is still independent of
of $\ell$, so that a universal bound such as (\ref{result}) exists.

The relation of the Hawking temperature to the quantum noise temperature
may of course only be a strange coincidence. After all, the minimal noise
temperature is only in the Schwarzschild case given exactly by
the Hawking temperature. Moreover, if one assumed that the QNMs
were already immersed in a bath with Hawking temperature, the
quantum measurement would add an additional noise.
However, for all black holes it seems that
a quantum noise temperature is associated with
the gravitational QNMs, and that the minimal noise temperature
is bounded from below by the Hawking temperature.  
Of course, also
the connection of the QNM frequency with area quantization and loop
quantum gravity could be spurious. 
But the possibility that there really exist such 
connections would be so exciting that it deserves further attention.
It could turn out, for example, that the entanglement of the (quantized) 
gravitational QNMs
with the black-hole quantum state gives rise to the
Bekenstein--Hawking entropy, through the mechanism of decoherence
\cite{deco}. This would immediately explain the universality of the entropy.

\medskip

\noindent
Discussions with Friedrich Hehl, Yuri Obukhov, and Carsten Weber
are gratefully acknowledged.

\end{document}